\documentstyle[11pt,iau_211,twoside]{article}
\input epsf

\def\edcomment#1{\iffalse\marginpar{\raggedright\sl#1\/}\else\relax\fi}
\reversemarginpar

\begin{document}
\title{Orbital Migration and the Brown Dwarf Desert}
 \author{Philip J. Armitage}
\affil{JILA, 440 UCB, University of Colorado, Boulder CO80309-0440, USA}
\author{Ian A. Bonnell}
\affil{School of Physics and Astronomy, University of St Andrews, North Haugh, Fife KY16 9SS, UK}

\begin{abstract}
The orbital elements of extreme mass ratio binaries will be modified by 
interactions with surrounding circumstellar disks. For brown dwarf 
companions to Solar-type stars the resulting orbital migration is sufficient 
to drive short period systems to merger, creating a brown dwarf 
desert at small separations. We highlight the similarities and the differences 
between the migration of brown dwarfs and massive extrasolar planets, and discuss 
how observations can test a migration model for the brown dwarf desert.
\end{abstract}

\section{Introduction}
Radial velocity surveys show that brown dwarfs are rarely found 
as close (within a few au) binary companions to Solar-type stars.
This article makes the case that the existence of this {\em brown 
dwarf desert} may have nothing to do with the formation of 
brown dwarfs. We argue instead that a brown dwarf desert is an expected 
consequence of the orbital migration of brown dwarfs within an evolving 
protostellar disk (Armitage \& Bonnell 2002). We make the basic 
assumption that brown dwarfs form in the same manner as stars -- during 
gravitational collapse -- and show that migration subsequently creates 
a desert by driving initially close star - brown dwarf binaries into mergers.

\section{Why migration matters for brown dwarfs}
A viscous accretion disk around a single star evolves such 
that the outward drift of a vanishingly small fraction of the 
mass absorbs all of the angular momentum, allowing the bulk of the 
disk to be accreted (e.g. Pringle 1981). The same reasoning 
applies to circumbinary disks (Pringle 1991). Given enough 
time, an arbitrarily small mass of gas, trapped outside the 
binary orbit, will soak up the binary orbital angular momentum, 
and drive the binary toward merger. 

Protoplanetary disks, however, have limited lifetimes of around 
1-10~Myr (Strom et al. 1989; Haisch, Lada \& Lada 2001). For 
{\em rapid} migration to occur, we require (roughly speaking) that 
the mass of gas in the disk be at least comparable to the mass of 
the embedded substellar object\footnote{This condition can be easily 
understood. If the angular momentum of the substellar object greatly 
exceeds that of the disk, it will take many viscous times for the 
disk to transport the `extra' angular momentum to large radii, and 
migration will be very slow.}. This requirement is almost certainly 
met for brown dwarfs that are binary companions to Solar-type stars, 
since both observations (Osterloh \& Beckwith 1995) and theoretical 
arguments (Lin \& Pringle 1990) suggest that the disk mass at 
early times comfortably exceeds typical brown dwarf masses. We 
therefore expect migration to be significant for brown dwarfs 
orbiting Solar mass stars, though not for brown dwarfs embedded 
in the weaker disks surrounding very low mass stars.

\section{Forming a brown dwarf desert}

\begin{figure}
\plotone{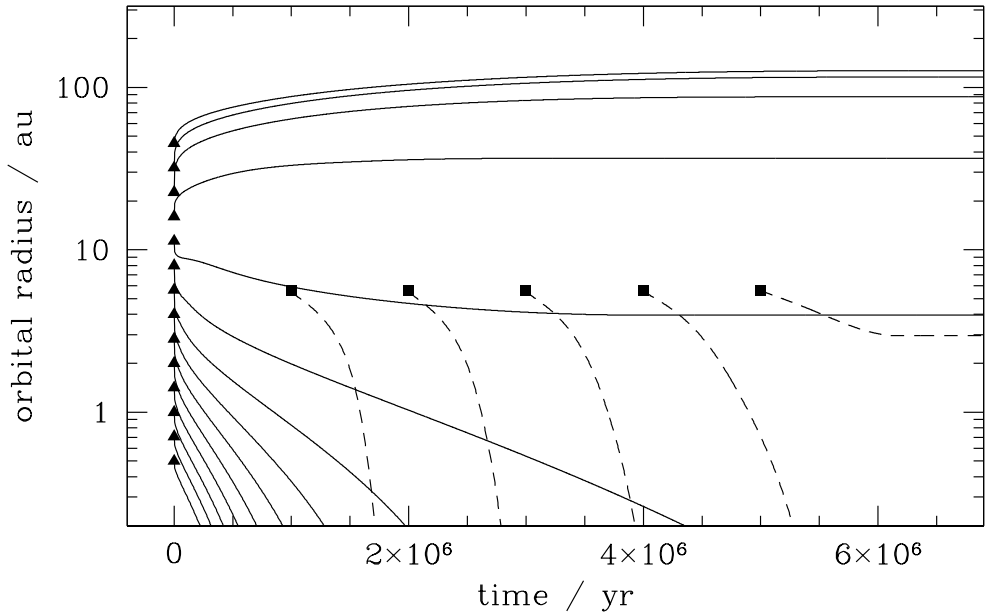}
\vspace{-2.0truein}
\caption{Comparison of the orbital evolution of $0.04 \ M_\odot$ brown dwarfs (solid 
         lines) and Jupiter mass planets (dashed lines) embedded within an evolving 
	 protoplanetary disk. Brown dwarfs form early, so to survive migration they 
	 must form within a narrow range of disk radii. Their evolutionary tracks 
	 diverge, implying that migration reduces the number 
	 of brown dwarfs at small orbital radii -- forming a brown dwarf desert.
	 Planets are even more susceptible to migration, but they can survive 
	 to be observed at radii of a few au by forming at a later epoch, when 
	 the disk is weaker and unable to drive significant migration.}	 
\end{figure}

To make quantitative predictions for the impact of migration on brown 
dwarfs, we have used a one dimensional model that was developed to 
study planetary migration (Armitage et al. 2002; see also Trilling 
et al. 1998; Trilling, Lunine \& Benz 2002). We solve for the coupled evolution 
of a viscous protoplanetary disk (using a variant of the disk model of Clarke, 
Gendrin \& Sotomayor 2001) and an embedded brown dwarf.
We assume that the brown dwarf orbit remains circular, not because 
we believe this to be true (Papaloizou, Nelson \& Masset 2001; Goldreich \& 
Sari 2002), but because of the difficulty of doing better. We also 
assume that the brown dwarf does not accrete further mass from the 
disk during migration. For circular orbits, this is likely to be a reasonable 
approximation for substellar objects with masses above a few Jupiter 
masses (Lubow, Seibert \& Artymowicz 1999).

Figure~1 shows the extent of the resulting migration for $0.04 \ M_\odot$ 
brown dwarfs embedded in disks whose initial mass is $0.1 \ M_\odot$. Brown 
dwarfs with initial orbital radii smaller than about 5~au are all swept 
in to small radii and probably merge with the star (conceivably, as has 
been suggested for extrasolar planets, some survive at very small 
radii as the brown dwarf equivalent of hot Jupiters). At larger 
radii, where the disk is expanding, outward migration 
occurs and pushes brown dwarfs out to radii of $\sim 10^2$~au. Between 
these extremes, brown dwarfs formed within a small window of initial 
orbital radii spread out to populate all short period orbits. This 
dilution of ${\rm d}N / {\rm d}\log a$ at small radii leads to a 
brown dwarf desert within this model.

We do not expect the desert -- if it results from migration -- to be 
completely devoid of brown dwarfs. A handful of brown dwarfs should 
end up in short period orbits having migrated inward from initially 
larger disk radii. Depending upon the assumed initial (pre-migration) 
distribution of brown dwarf orbital radii, we have estimated that the 
frequency of brown dwarfs within 4~au should 
be depleted by a factor of 5-10 (Armitage \& Bonnell 2002).

Why is there a brown dwarf desert at the same radii where there is an 
abundance of massive extrasolar planets (Marcy \& Butler 2000)? The 
arguments given above suggest -- correctly -- that the lower masses 
of extrasolar planets should make them even more vulnerable to migration 
than brown dwarfs. Massive planets, however, probably do not form via 
the same mechanism as brown dwarfs (though for contrary opinions see 
Boss 1998; Armitage \& Hansen 1999). Conventionally, they form instead 
via core accretion, a slow process that may take several Myr, especially 
in the outer parts of the disk. As shown in Figure~1, this allows planets 
to escape destruction via mergers by forming at a later epoch, 
when the protoplanetary disk is of lower mass and unable to force 
the planets into mergers.

\section{Observational tests}
We have argued that one should not conclude, based solely on the existence 
of the brown dwarf desert, that brown dwarfs cannot form as close 
binary companions to Solar-type stars. Migration at a later epoch 
can accomplish the same end result, and makes several additional predictions.
\begin{itemize}
\item[(i)]
Migration occurs if the binary mass ratio $q = M_{\rm BD} / M_*$ is 
comparable to or less than the disk mass ratio $M_{\rm disk} / M_*$. 
This is unlikely to be true for brown 
dwarfs around very low mass stars. Therefore, no brown 
dwarf desert is expected around M stars. 
\item[(ii)]
The same argument applies to M star companions to stars of a few Solar 
masses -- {\em if} the latter have disks that live for several viscous 
times. Migration could clear out a low mass stellar desert at small separations around 
stars modestly more massive than the Sun.
\item[(iii)]
Migration takes of the order of a Myr. 
No brown dwarf desert is expected around the youngest pre-main-sequence 
stars.
\end{itemize}
We emphasize that the first two predictions are not unique to this model.
Accretion onto forming binary systems also abhors extreme mass ratios 
(Bate 2000), and provides an alternative theoretical explanation for 
the brown dwarf desert. The third prediction is more discriminatory.
Detection of a higher frequency of brown 
dwarfs around young pre-main-sequence stars would provide strong support 
for the model presented here. Competing theories based on differential 
accretion (Bate 2000) and dynamical interactions (Reipurth \& Clarke 2001) 
predict that the desert is established on a dynamical timescale, which 
is at least an order of magnitude shorter than the timescale on which 
migration occurs.

\end{document}